\documentclass{aa}
\usepackage{amsmath}
\usepackage{graphicx}
\usepackage{natbib}
\bibpunct{(}{)}{;}{a}{}{,} 
\usepackage{txfonts}
\usepackage{color}
\begin{document}

   \title{
A stochastic simulation of the propagation of Galactic cosmic rays reflecting the discreteness of cosmic ray sources.\\  Age and path length distribution.
}

   \subtitle{}
   
   \titlerunning{Propagation of GCRs reflecting the discreteness of CR sources}

   \author{S. Miyake
          \inst{1}
          \and
          H. Muraishi
          \inst{2}
          \and
          S. Yanagita\inst{3}
          }

   \institute{Department of Electrical and Electronic Systems Engineering, 
                Ibaraki National College of Technology, Ibaraki 312-8508, Japan\\
              \email{miyakesk@ee.ibaraki-ct.ac.jp}
         \and
             School of Allied Health Sciences, Kitasato University,
             Kanagawa 252-0373, Japan\\
         \email{muraishi@ahs.kitasato-u.ac.jp}
         \and
             Faculty of Sciences, Ibaraki University,
             Ibaraki 310-8512, Japan\\
         \email{yanagita@mx.ibaraki.ac.jp}
             }

   \date{Received September 15, 1996; accepted March 16, 1997}

 
  \abstract
   {}
  {The path length distribution of Galactic cosmic rays (GCRs) is the fundamental ingredient for modeling the propagation process of GCRs based on the so-called weighted slab method.
  We try to  derive  this distribution numerically by taking into account  the discreteness in both space and time of occurrences of supernova explosions where GCRs are suspected to be born.
The resultant age distribution and ratio of B/C are to be compared with recent observations.
}
   {  
We solve numerically the stochastic differential equations equivalent to the Parker diffusion-convection equation which describes the propagation process of GCR in the Galaxy.
We assume the three-dimensional diffusion is an isotropic one without any free escape boundaries.
We ignore any energy change of GCRs and the existence of the Galactic wind for simplicity. 
We also assume axisymmetric configurations for the density distributions of the interstellar matter and for the surface density of supernovae.
We have calculated age and path length of GCR protons arriving at the solar system with this stochastic method.
The obtained age is not the escape time of GCRs from the Galaxy as usually assumed, but the time spent by GCRs during their journey to the solar system from the supernova remnants where they were born.
 }
   { 
The derived age and path length show a distribution spread in a wide range even for GCR protons arriving at the solar system with the same energy.
The distributions show a cut-off at a lower range in age or path length depending on the energy of GCRs.
These cut-offs clearly come from the discreteness of occurrence of supernovae.
The mean age of GeV particles obtained from the distributions is consistent with the age obtained by direct observation of radioactive secondary nuclei.
The energy dependence of the B/C ratio estimated with the path length distribution reproduces reliably the energy dependence of B/C obtained by recent observations in space.
}
   {}


   \maketitle
%

\section{Introduction}

Supernova remnants (SNRs) are believed to be the sources of Galactic cosmic rays (GCRs) from the argument of energetics and the diffusive shock acceleration mechanism of particles expected in the shocks associated with the remnants (e.g., Blandford $\&$ Eichler \citeyear{blandford}).
This view is corroborated by recent TeV gamma-ray observations of SNRs by CANGAROO (e.g., Muraishi et al. \citeyear{muraishi2000}), HESS measurement (e.g.,  Aharonian et al. \citeyear{aharonian}), and the other experiments.
Supernovae (SNe) are believed to occur once per 30 to 100 years in our Galaxy.
Hence we must take into account  the discreteness of SNe in space and time in the investigation of the propagation process of GCRs from parent SNRs to the solar system.
However, the numerical codes widely used to calculate the propagation of the GCRs in the Galaxy such as GALPROP (e.g., Strong $\&$ Moskalenko \citeyear{strong1998}) and DRAGON (e.g., Evoli et al. \citeyear{dragon}) may not necessarily be suitable to fully include  this discreteness in the calculation of propagation of GCRs.
The important effects of this discreteness imprinted in the nature of GCRs observed in the solar system have been pointed out by various authors (Higdon $\&$ Lingenfelter \citeyear{higdon}; Taillet et al. \citeyear{taillet}; Mertsch \citeyear{mertsch}; Blasi $\&$ Amato \citeyear{blasi}; Bernard et al. \citeyear{bernard}).
However, this is the first time that simulations of the propagation of GCRs that fully take into account  the stochasticity coming from the spatial and temporal discreteness of occurrences of SN explosion have  been attempted.

In this paper, we propose a new and fully three-dimensional stochastic numerical method to calculate age distribution and path length distribution (PLD) of GCRs reflecting the discreteness of the occurrence of supernova explosions. 
Our stochastic method is based on the equivalence of a coupled set of stochastic differential equations (SDEs) to the Parker convection-diffusion equation describing the propagation of GCRs.
The power of this SDE method has been demonstrated in the study of the solar modulation phenomena where the method revealed some important physical properties, which are difficult -- if not impossible -- to obtain by other numerical methods such as the distribution of the arrival time from the heliospheric boundary to the earth and the distribution of energy loss during this period (Yamada et al. \citeyear{yamada}; Zhang \citeyear{zhang}).
Similarly, it is expected that our method based on SDEs  will reveal new aspects of the propagation process of GCRs.
In this study, we assume that three-dimensional diffusive propagations of GCR are isotropic and we do not impose free escape boundaries.  
We also assume axisymmetric configurations for the density distributions of the interstellar matter and for the surface density of supernovae.
We present various results of simulations with SDEs including the energy dependence of the B/C ratio to be compared with recent observations by CREAM (Ahn et al. \citeyear{ahn}){\bf ,} AMS-02 (Oliva et al. 2013\nocite{AMS02}), and PAMELA (Adriani et al. \citeyear{PAMELA}).

\section{Models and numerical simulation}

In this study we numerically investigate the propagation of GCR protons which  originated in SNRs in the Galaxy.
We assume that the acceleration of GCRs in SNRs continues uniformly in time up to $10^5$ years after the explosion of the parent supernova, although it has been suggested in recent studies that the escape time of GCRs from SNRs depends on their energy (e.g., Ohira et al. \citeyear{ohira}; Caprioli et al. \citeyear{caprioli}; Drury \citeyear{drury}).
The radius of SNRs at the age of $10^5$ years is assumed to be 30 pc by following the Sedov model where the values of $10^{51}~\mathrm{ergs}$, $10^9~\mathrm{cm}~ \mathrm{sec}^{-1}$, and $1~\mathrm{proton}~\mathrm{cm}^{-3}$ for the total explosion energy, the velocity of the shock wave, and the ambient matter density, respectively, are adopted.
We also assume that the supernova (SN) occurs randomly both in time and in space within the radius of $20~\mathrm{kpc}$ from the center of the Galaxy.
The frequency of occurrences of SN is assumed to be 3 times in 100 years in the Galaxy.
The surface density of the SN rate has a galactocentric radial dependence scaled to the molecular gas given by  Williams $\&$ McKee (\citeyear{williams}), and has a Gaussian height distribution $P_{SN}=(2\pi {\alpha}^2_{SN})^{-1/2}~\mathrm{exp} \{ -z^2/(2{\alpha}^2_{SN})\}~\mathrm{kpc}^{-1},$ where $\alpha_{SN}=0.070~\mathrm{kpc}$ (Boulares $\&$ Cox \citeyear{boulares}).

The transport and distribution of GCRs is described by the Fokker-Planck equation (FPE) also called the Parker convection-diffusion equation,
\begin{equation}
\begin{array}{llllll}
\dfrac{\partial f }{\partial t} &=&  \boldsymbol{\nabla} \cdot 
        \left( {\boldsymbol{\kappa}}  \cdot   \boldsymbol{\nabla} - {\bf{V}} \right) f + 
        \dfrac{1}{3} \left( \boldsymbol{\nabla} \cdot {\bf{V}} \right) 
        \dfrac{1}{p^2} \dfrac{\partial}{\partial p} \left(p^3 f\right) \\
        & & +  \dfrac{\partial}{\partial p}\left( {\dot p_c} f \right)   +  \dfrac{1}{p^2}  \dfrac{\partial}{\partial p} \left(D_p p^2 \dfrac{\partial}{\partial p} f\right)~,
\end{array}
\label{eq1}
\end{equation}
where $ f \left({\bf r}, {\bf p}, t\right)$ is the cosmic-ray distribution function, $\bf{r}$ and $\bf {p}$ indicate the coordinate values of a generic point in the phase space, $t$ is the time, $\bf{V}$ is the velocity of the Galactic wind, ${\boldsymbol{\kappa}}$ is the spatial diffusion coefficient tensor, $D_p$ is the diffusion coefficient in the momentum space, and  ${\dot p_c} \equiv d p_c / dt$ is the momentum gain or loss rate. 
The full set of SDEs equivalent to the FPE (\ref{eq1}) is written as
\begin{equation}
 \begin{array}{llllll}
d{\bf{r}} & = & \left( \boldsymbol{\nabla} \cdot {\boldsymbol{\kappa}}
  - {\bf{V}} \right)dt 
  + \sum_{s} {\boldsymbol{\sigma}}_{s} dW_{s}(t)~, \\
   & &\\
dp & = & \left\{ \dfrac{2 D_p}{p}+ \dfrac{\partial D_p}{\partial p} -\dfrac{1}{3} p\left( \boldsymbol{\nabla} \cdot {\bf{V}} \right) +{\dot p_{c}} \right\} dt+\sqrt{2D_p} dW_{p}(t)~,
  \end{array}
\label{eq2}
\end{equation}
where $\bf{r}$ and $p$ indicate the position and the momentum of pseudo-particle respectively (Hereafter we refer to "pseudo-particle" as just "particle" if needed.), $\sum_{s} {\sigma}^{\mu}_{s} {\sigma}^{\nu}_{s} = 2 {\kappa}^{\mu \nu}$, and $dW_{s}$ and $dW_{p}$ are the Wiener processes given by the Gaussian distribution.
In this study, we assume an isotropic three-dimensional diffusive propagation without imposing free escape boundaries and neglect any energy changes and the existence of the Galactic wind for simplicity.
Then, the full set of SDEs (\ref{eq2}) is reduced to a simple SDE
\begin{equation}
  d{\bf{r}}=\sum_{s} {\boldsymbol{\sigma}}_{s} dW_{s}(t)~.
  \label{eq3}
\end{equation}
We solved this SDE backward in time (see Yamada et al. \citeyear{yamada}; Zhang \citeyear{zhang}).
In other word, we follow the sample path of the particles (protons) with various fixed energies at the earth backward in time until the particle arrives at some active SNR.

If we assume an isotropic diffusion, the diffusion tensor $ {\kappa}^{\mu \nu}$ is reduced to a scalar $\kappa$ represented as $\frac{1}{3} l_{m.f.p.} v$, where $l_{m.f.p.}$ and $v$ are the diffusion mean free path and the velocity.
The mean free path $l_{m.f.p.}$ is represented as $a \{ p/(1 \mathrm{GeV/c}) \}^{2-\delta}$ with two parameters $a$ and $\delta$.
We adopted the value $( a~ \left[ \mathrm{cm}\right], \delta )$ as $(4.1 \times 10^{17},  \frac{5}{3} )$, $( 3.6 \times 10^{17}, \frac{3}{2} )$, and $( 2.8 \times 10^{17}, 1 )$ for the Kolmogorov-type,  the Kraichnan-type, and the Bohm-type models, respectively.
The value of $a$ for each model is determined so that our result of the B/C ratio at 1 GeV coincides with the observed value by AMS-02 as shown later.
This normalization may be allowed because the effect of the solar modulation on the ratio of B/C is estimated less than $6\%$ with reasonable values for the modulation parameter according to the force field theory.

The path length $X$ is calculated simultaneously with the SDE as $dX=\rho v dt$ where $\rho$ is the density of the interstellar matter.
The density $\rho$ is calculated with the model of interstellar medium adopted by Higdon $\&$ Lingenfelter (\citeyear{higdon}). 
In this model, the interstellar mediums consist of five major components, the molecular gas, cold neutral gas, warm neutral gas, warm ionized gas, and hot tenuous medium.
The density of the molecular gas and the other components are given by  Williams $\&$ McKee (\citeyear{williams}) and Ferri$\Grave{\mathrm{e}}$re (\citeyear{ferriere}) respectively, as a function of the Galactocentric radius and the height from the Galactic plane.

In our algorithm, first we generate a list of SNe which might have occurred during the last $5\times10^8$ yr in our Galaxy so that their spatial position ${\bf r_{SN}}$ and occurrence time $t_{SN}$ are randomly distributed  to be consistent with the conditions as stated before.
Next we start to follow the sample path ${\bf r}(t)$ with certain energy by the SDE with an appropriate time step $\Delta t$.
The path length for each time step can be calculated by $v \rho \Delta t$.
Sample paths start at the solar system which is located at $\left(8.5~\mathrm{kpc},0~\mathrm{kpc},0~\mathrm{kpc} \right)$ for a galactocentric Cartesian coordinate system $\left( x, y, z \right)$, and then run backward in time until they hit some active SNR centered at the parent SN on the prepared list where the acceleration of CRs is still going on.
If the sample path ${\bf r}(t)$ at some time $t$ get into some SNR, in other words ${\bf r}(t)$ satisfies $|{\bf r}(t)-{\bf r_{SN}}| \le 30~\mathrm{pc}$, $t_{SN}-10^5~\mathrm{yr} \le t \le t_{SN}$, then we stop to calculate the sample path and we record the position, the arrival time, and the total path length, and then we restart the same procedure again for another particle.
We follow the sample path up to $5\times 10^8$ yr while the particle fails to hit some active SNR.
We adopt this figure of $5\times 10^8~\mathrm{yr}$ so that it should be long enough compared with the age of GCR $\sim 15~\mathrm{Myr}$ estimated in the frame of the leaky box model based on direct observations of radioactive secondary nuclei in the energy range of a few 100 MeV/n by ACE (Yanasak et al. \citeyear{yanasak}). 
The adequacy of this figure as the time limit is   shown later by our simulation results.
By this numerical experiment, we can simultaneously obtain the age distribution and the PLD for particles observed with various energies at the earth.
We want to emphasize that the age of GCRs obtained by this method is not the escape time from the Galaxy but the time spent by the particles during their journey to the earth from the SNR where they were born\footnote{After we submitted our paper, Lipari (\citeyear{lipari})  presented an interesting article in arXiv where he develops the concept of cosmic ray ages that is similar to ours.}.
Accordingly, the value of this age should depend on the location of observers in our Galaxy.

\section{Results and discussions}

\begin{figure}[t]
   \centering
      \includegraphics[width=8cm]{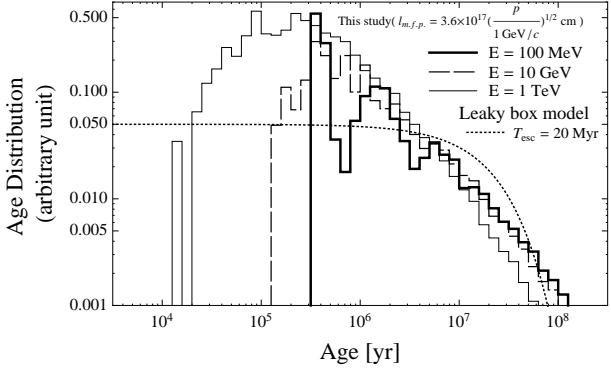}
      \caption{Calculated age distribution of GCR protons with some fixed energies in the solar system for the case of $\delta=\frac{3}{2}$. The thick line, the long-dashed line, and the thin line indicate the age distribution of protons which arrive at the solar system with a fixed energy of 100 MeV, 10 GeV, and 1 TeV, respectively. The dotted line indicates the age distribution expected by the leaky box model with the escape lifetime of 20 Myr. 
              }
         \label{fig1}
   \end{figure}

\begin{figure}[t]
   \centering
    \includegraphics[width=8cm]{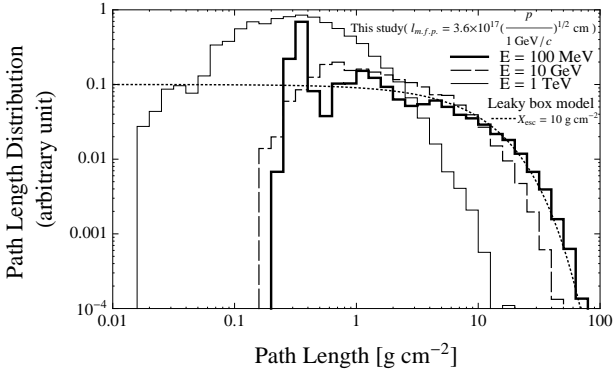}
      \caption{Calculated PLD of GCR protons with some fixed energies in the solar system for the case of $\delta=\frac{3}{2}$. For an explanation see the caption to Fig. \ref{fig1}. The dotted line indicates the PLD expected by the leaky box model with the escape path length of $10~\mathrm{g}~ \mathrm{cm}^{-2}$.
              }
         \label{fig2}
   \end{figure}

The age distribution and the PLD of protons are shown in Fig. \ref{fig1} and Fig. \ref{fig2}, respectively, for some fixed energies in the solar system overlaid with the expected distributions by the leaky box model.
It can be seen that the resultant age and the path length are distributed over a significantly wide range depending on the observed energy even for particles with the same observed energy.
These distributions for each energy are certainly not obtained directly by observations; however, they play a crucial role in determining the model of propagation process of GCRs by the weighted slab method.
It may be difficult, if not impossible, to estimate these distributions by other methods than ours.
As we stated before, it may be recognized that our choice of $5 \times 10^8~\mathrm{yr}$ as the time limit to follow sample paths is justified by the resultant distributions shown in Fig. \ref{fig1} where almost all of sample paths are distributed in the range lower than $10^8~\mathrm{yr}$ even for the lowest energy particles of 100 MeV, which have been revealed to have the longest mean age within a few representative energies shown in Fig. \ref{fig1}.

It is remarkable that a rapid decrease or a cut-off can be seen in the distributions shown in Fig. \ref{fig1} and Fig. \ref{fig2} below a certain age or path length depending on the energy of particles.
Obviously no such cuts can be seen for the leaky box model.
The origin of the cuts is traced back to the discreteness of the occurrence of SN as the source of GCRs.
The position of the cut-off shifts to a lower range when the energy of the particle becomes higher.
These tendencies arise because  the diffusion time of GCRs from their birth place to the solar system becomes shorter when the energy of the particle becomes high.
The effects of the discreteness of the GCR sources on the age distribution and the PLD have been investigated by the myriad-source model by Higdon $\&$ Lingenfelter (\citeyear{higdon}) and Taillet et al. (\citeyear{taillet}); however, they failed to discover the existence of the cuts in the distributions because they averaged out the spatial position of SNe.
The existence of cuts in the PLD was pointed out by Cowsik $\&$ Wilson (\citeyear{cowsik}) from quite a different argument of the nested leaky box model.
Garcia-Munoz et al. (\citeyear{garcia}) suggested a lack in short path lengths in the PLD from a comparison of the B/C ratio to the sub-Fe/Fe ratio.
Our model succeeded in predicting the missing short path lengths quite naturally and quantitatively.

  \begin{figure}[t]
   \centering
   \includegraphics[width=8cm]{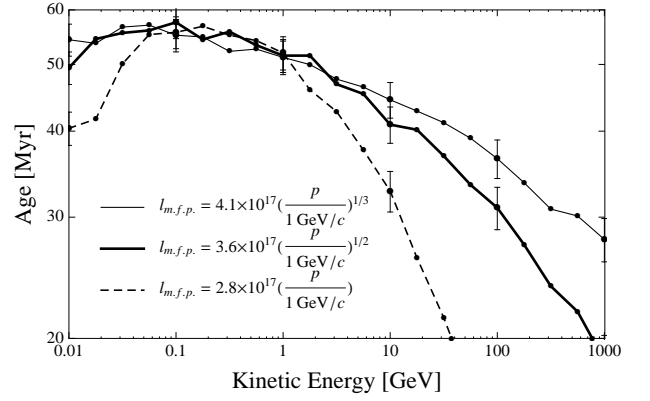}
   \caption{Mean values of the age distribution of GCR protons. The thin line, the thick line, and the dashed line indicate the mean values for the case of $\delta=\frac{5}{3}$, $\frac{3}{2}$, and 1, respectively. The error bars shown for some points indicate typical statistical errors of $\pm 3 \sigma$ in our simulation.
              }
         \label{fig3}
   \end{figure}
     
    \begin{figure}[t]
   \centering
   \includegraphics[width=8cm]{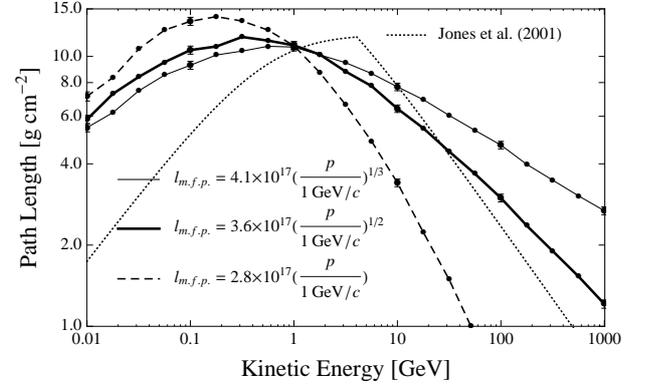}
      \caption{Mean values of the PLD of GCR protons.
The dotted line indicates the mean values of the PLD assumed by Jones et al. (\citeyear{jones}).
The other line symbols are as in Fig. \ref{fig3}.
              }
         \label{fig4}
   \end{figure}

Both distributions shown in Fig. \ref{fig1} and Fig. \ref{fig2} for 100 MeV particles exhibit clear peaks near the cut-off regions.
We have checked that the peaks are due to particles that originated in several relatively young nearby SNRs.
As mentioned before, it is unfortunate that these remarkable features cannot be checked directly by observations.
However, the effects may be reflected indirectly in observed ratios of secondary to primary nuclei especially for unstable secondary nuclei.
Clear peaks are not seen in the distributions for 10 GeV and 1 TeV particles.
This is understandable from the fact that the fraction of the particles from the same nearby SNRs that make the peaks for 100 MeV particles becomes smaller for the higher energy particles because the number of SNRs where these particles were born becomes higher.
This is caused because the value of the diffusion coefficient increases with the particle energy.
   
We  note that the ages obtained by observations for GCRs with certain energies may be the values averaged over distributions similar to those shown in Fig. \ref{fig1}. 
Calculated average ages to be compared with observations are shown in Fig. \ref{fig3} for three cases with different forms of the diffusion mean free path. 
As anticipated, the ages decrease monotonically with energies above about 1 GeV. 
The mean value of $\sim 5\times10^7~\mathrm{yr}$ around at 1 GeV agrees well with the age obtained by the observation of radioactive secondary nuclei (e.g., Yanasak et al. \citeyear{yanasak}).

Similarly, mean values of PLD as usual notion are calculated from such distributions as shown in Fig. \ref{fig2} for each energy.
The results are shown in Fig. \ref{fig4} for the same three cases as are shown in Fig. \ref{fig3}. 
One of the notable features in the PLD is that a peak value is attained at a few 100 MeV. 
Many authors have assumed that the PLD is piecewise continuous as a function of energy with a peak at around a few GeV to explain phenomenologically the observed energy dependence of secondary to primary ratios in GCR (e.g., Jones et al. \citeyear{jones}; Ptuskin \citeyear{ptuskin2012}).
As an example we overlaid  the PLD by Jones et al. in Fig. \ref{fig4}.
Our model reproduces reliably, though qualitatively, the PLD with such a form without introducing a break point by hand. 
The obtained PLD decreases monotonically with energies  around 1 GeV depending on the form of the diffusion mean free path, as was anticipated. 

\begin{figure}[t]
   \centering
   \includegraphics[width=8cm]{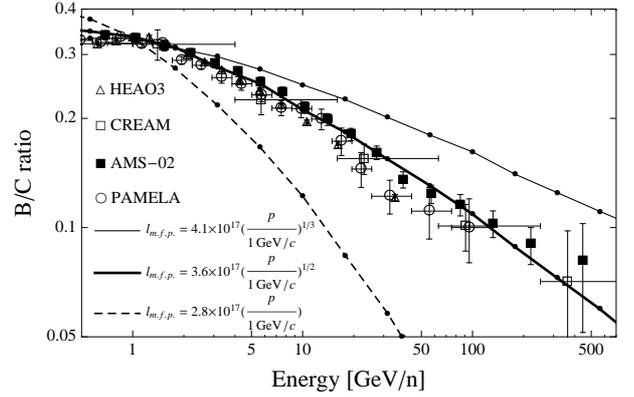}
      \caption{Derived B/C ratios for three cases with different forms of the diffusion mean free path overlaid with the observational data by HEAO3, CREAM (compilation by Maurin et al. \citeyear{maurin} and the references cited therein), AMS-02 (Oliva et al. 2013), and PAMELA (Adriani et al. \citeyear{PAMELA}).
The line symbols are as in Fig. \ref{fig3}.
              }
         \label{fig5}
   \end{figure}

Once we get the PLDs for protons with arbitrary energies, we can calculate  abundance ratios of the secondary to primary elements at corresponding energies by the weighted slab method if we are allowed to assume the PLDs for the heavy elements are the same as those  of protons. 
In the calculations, we should use the PLDs shown in Fig. \ref{fig2} prepared in advance for each energy and should not use the averaged PLD shown in Fig. \ref{fig4} because, as mentioned earlier, the PL is distributed in a wide range even for particles with the same energy.
If we use the averaged PLD as is usually done,  there are subtle differences  in the ratios even though we do not show the details here. 
We present the calculated energy dependence of B/C ratios in Fig. \ref{fig5} for the same three different forms of the mean free path as shown in Fig. \ref{fig3} and Fig. \ref{fig4} together with recent observational results.
Here we calculated the B/C ratio in the energy range where the effect of solar modulation is estimated to be small by the force field theory.
In our calculation, only C, N, and O are assumed to be the parent nuclei of B. 
The source abundance of each parent nuclei was taken from Strong $\&$ Moskalenko (\citeyear{strong2001}). 
Relevant nuclear data were referred to Garcia-Munoz et al. (\citeyear{garcia}), Webber et al. (\citeyear{webber2003}), and Ramaty et al. (\citeyear{ramaty}). 

The numerical values multiplied to each form of the diffusion mean free paths presented in Fig. \ref{fig5} were chosen so as to reproduce best the value of B/C at 1 GeV/n obtained by AMS-02 as mentioned earlier. 
As clearly seen, the observational results are most reliably reproduced in the whole range of energy when the power law index of momentum dependence is 1/2. 
This result indicates that the value of $\delta$ is 3/2 and supports the Kraichnan model for the interstellar turbulence.

We did not address here discussions on the B/C in the energy range below $\sim 1~\mathrm{GeV/n}$ where the ratio may be influenced to somewhat high degrees by the convection from the Galactic wind (e.g., Strong $\&$ Moskalenko \citeyear{strong1998}), the reacceleration in the interstellar medium (e.g., Simon et al. \citeyear{simon}; Giler et al. \citeyear{giler}; Seo $\&$ Ptuskin \citeyear{seo}), and the various plasma processes reflected in the diffusion coefficient (e.g., Shalchi et al. \citeyear{shalchi}; Evoli $\&$ Yan \citeyear{evoli2014}). 

So far we have only discussed   nuclear components in the GCR; however, we should not  overlook the effects of the discreteness of SNe occurrences on electron components in the GCR.
Kobayashi et al. (\citeyear{kobayashi}) pointed out the possible peculiar energy spectrum of electrons in the TeV energy range where electrons  from only a few nearby SNRs could contribute because of severe energy losses suffered by electrons from other distant SNRs. 
These interesting features will be revealed by scheduled experiments such as CALET (Akaike et al. \citeyear{akaike}) in the very near future.

\section{Summary}

We demonstrated a new numerical method to investigate the propagation process of GCRs in our Galaxy which is based on the equivalence of the diffusion convection equation to a coupled stochastic differential equations. 
Assuming SNRs are the birthplace of the GCRs, we examined the effects on the nature of the GCRs arriving at the solar system caused by spatial and temporal discreteness of the SN occurrences in a fully 3D scheme though for a simplified model of an isotropic diffusion. 

The resultant PLD for particles arriving at the solar system with the same energy exhibits a distribution spread over a wide range of path length values with a sharp cut below some lower path length depending on the energy. 
The existence of PLDs with a cut has been anticipated by studies of the abundance ratios of secondary to primary elements. 
The cut is clearly brought about by the discrete occurrences of SN explosions and accordingly may not be attained by models with continuous source distributions as usually assumed. 
The age distribution obtained simultaneously with PLDs exhibits a characteristic distribution with a cut also caused  by the discreteness of SN explosions. 
The calculated average age agrees well with the value obtained by observations of radioactive secondary nuclei. 

The energy dependence of B/C ratios from the obtained PLDs was calculated by the weighted slab method for a few forms of the diffusion mean free path as a function of momentum. 
It was found that the model with the mean free path proportional to the square root of particle momentum reproduces best the recent observational results.
This finding supports the idea of the Kraichnan  model for the interstellar turbulence.

Results by simulations which take into account important factors not considered in this work such as the Galactic wind, various energy changes including reaccelerations, and the more detailed global structure of the Galaxy with the magnetic fields will be presented in another paper in the near future.

\begin{acknowledgements}
We thank J. Nishimura, T. Kamae, T. Shibata and S. Torii for their invaluable comments for this work.
We are grateful to the referee for critical comments and useful suggestions.
This work is supported by JSPS KAKENHI Grant Number 26800145 (S.M.) and 22540264 (S.Y.).
\end{acknowledgements}


\bibliographystyle{aa}
\bibliography{bib-pld}


\end{document}